\documentclass[11pt,a4paper]{article}
\usepackage{graphicx}
\renewcommand{\dj}{\hbox{d\hskip-1.1ex{\raise0.640ex\hbox{--}}\skip 0.70ex}}

\newcommand{\calL}{\mathcal{L}}

\newcommand{\lsim}[1]{
\setlength{\unitlength}{12pt}
\begin{picture}(1.4,1.)
\put(.7,-0.3){\makebox(0.0,1.)[t]{$<$}}
\put(.7,-0.3){\makebox(0.0,1.)[b]{$\sim$}}
\end{picture}#1}

\begin{document}

\thispagestyle{empty}
\begin{flushright}
ZTF--03/02
\end{flushright}

%\end{flushleft}

%\vspace{1.0cm}

\begin{center}

  \begin{Large}

  \begin{bf}

On the instanton--induced portion of the nucleon strangeness II:
the MIT model beyond the linearized approximation

  \end{bf}

  \end{Large}

\vspace{1cm}

  \begin{Large}

\center{D. Klabu\v{c}ar$^{1,\rm a}$, K. Kumeri\v{c}ki$^{1,\rm b}$, 
D. Mekterovi\'{c}$^{2,\rm c}$, \\ B. Podobnik$^{1,\rm d}$ }

\vspace{1.2cm}

{\footnotesize 

{\small $^1$ Department of Physics, Faculty of Science, University
of Zagreb, \\ Bijeni\v cka c. 32, HR-10000 Zagreb, Croatia    }

\vspace{0.8cm}

$^2$ Rudjer Bo\v{s}kovi\'{c} Institute,
         P.O. Box 180, HR-10002 Zagreb, Croatia \\ }

\vspace{0.7cm}

\end{Large}

\end{center}

\vspace{0.1cm}
\begin{center}
  {\bf Abstract}
\end{center}
\begin{quotation}
\noindent
	
\noindent 
If instantons are introduced into the MIT bag model 
in such a way that bag radii are allowed to vary, 
the MIT bag interior can accommodate instanton 
density which is by an order of magnitude larger 
than in the case when the radii are fixed (although 
it is still significantly smaller than in the 
nonperturbative QCD vacuum). The instanton contribution 
to baryon mass shifts is also correspondingly larger.
The instanton-induced part of the scalar strangeness 
of the nucleon MIT bag is an order of magnitude larger
than found previously, within the linearized approximation.
The decrease of the model radii (which is associated 
with the increase of the instanton density)
is very favorable from the standpoint of nuclear physics.

\vspace{0.5cm}


\end{quotation}                 

\vfill

----------------------------------------

$^{\rm a}$ e-mail: klabucar@phy.hr

$^{\rm b}$ e-mail: kkumer@phy.hr

$^{\rm c}$ e-mail: dmekter@rudjer.irb.hr

$^{\rm d}$ e-mail: bp@phy.hr. Present address: Faculty of Civil Engineering,\\
             University of Rijeka, HR-51000 Rijeka, Croatia

\newpage                                        

\markright{$$\small 
On the instanton--induced portion of the nucleon strangeness II: ...
$$}

%\Large

\section{Introduction}
\label{INTRO}

Characteristics of nonperturbative QCD make intractable many 
calculations at low and intermediate energies. Effective quark 
models therefore retain their usefulness in numerous applications.
For example, ref. \cite{Klabucar:1998pb} 
used  the instanton-extended version \cite{2}{\footnote{In ref. \cite{2}, 
incorporating instanton-induced 
interactions into the original MIT bag model \cite{3,6}, 
was inspired by   
an analysis \cite{1} made in the constituent quark model,
which found that an effective instanton interaction led to 
as satisfactory a description of mass splittings of baryons
as the conventional approach using one-gluon exchange.}}
of the MIT 
bag model \cite{3,6} in one of many studies of strangeness 
in nucleons \cite{Alberico:2001sd}. 
However, the approach of ref. \cite{2} contained the so-called
linearized approximation, amounting to freezing the
baryon radii in their original MIT values. In the present
paper we remove this approximation and calculate the effects 
thereof on the baryon mass splittings, and also on the 
nucleon strangeness results of ref. \cite{Klabucar:1998pb}. 
We also explore whether this enables the resolution or 
alleviation of a long-standing inconsistency between the MIT 
bag model and nuclear physics: the standard nuclear physics 
descriptions employ independent nucleons, while the nucleon 
MIT bag radius is too large for that \cite{GerryEtAl}.

The first approach to consider the instanton-induced interaction
within a bag model was due to Kochelev \cite{Kochelev:de}. It is 
nevertheless important to note that he considered his own bag model 
\cite{Kochelev:et}, which is somewhat different from the MIT one. 
As explained in detail in ref. \cite{2}, that line of research 
\cite{Kochelev:et,Kochelev:de,DorokhovKochelev,Dorokhov:hk,Dorokhov:ym,Dorokhov:1994fv,Kochelev:1997ux} 
is therefore rather different from the one in ref. \cite{2} and 
in the present paper, where we stay as close as possible to the
original MIT bag model \cite{3,6}. The only modification with
respect to the MIT model is
the inclusion of the instanton-induced interaction \cite{2}. 
This inclusion is necessary also {\it inside} the MIT bag if one
allows for the nonvanishing (even if small) probability
of penetration of the instanton liquid from the surrounding
nonperturbative QCD vacuum, into the ``perturbative'' MIT
bag interior. The instanton density $n$ used in this effective
instanton-induced interaction inside the bag, is of course 
reduced with respect to the density in the nonperturbative 
QCD vacuum: the smaller the probability of this penetration,
the larger the reduction. The reduced value of $n$ appropriate 
for the MIT bag interior comes out as a result of our model 
calculation and fitting.

Besides defining how to incorporate instantons 
in the MIT bag model and finding the baryon mass shifts caused 
by the effective instanton-induced quark-quark interaction, 
ref. \cite{2} explored the modification of the 
too large value (required by 
the ``instantonless" model fits, {\it e.g.}, \cite{6}) 
of the strong coupling constant $\alpha_{c}$ 
used in the supposedly perturbative MIT bag interior. 
The conclusion of ref. \cite{2} was 
that the change in $\alpha_{c}$ was in the desired direction, 
{\em i.e.}, it was reduced, 
but only by about $6 \%$, which was insufficient to achieve 
improvement in the 
consistency of the perturbative description of the bag interior. 
Actually, it turned out that instanton effects should 
generally be small in the MIT bag model, since the 
instanton (plus anti-instanton) density $n$ appropriate 
for the MIT bag interior, was found \cite{2} to be very much depleted  
with respect to the instanton density estimated ({\it e.g.}, by 
refs. \cite{5,8,9,4,Schafer:1996wv})
for the true, nonperturbative QCD vacuum. 
The corresponding instanton-induced mass shifts were of the order 
of only a few MeV. 
However, the analysis of ref. \cite{2} 
(and, consequently, of ref. \cite{Klabucar:1998pb})
contained an important 
simplifying assumption: it kept the baryon radii ``frozen" 
in the values obtained by DeGrand {\it et al.} \cite{6}. 
In this way, a full refitting of the bag 
model parameters (now also including the instanton density $n$
inside the bag) was avoided. In fact, this highly nonlinear problem 
was thereby reduced to solving the four linear equations for the 
four adjustable model parameters that enter the energy functional 
linearly: $\alpha_{c}$, the volume-energy density $B$, 
the zero-point energy parameter 
$Z_{0}$, and a parameter  new to the MIT bag model, 
namely, the instanton density  $n$. 
(The quark masses were also not allowed to vary. 
Ref. \cite{2} adopted the quark masses of 
DeGrand {\it et al.} \cite{6} in order to be able 
to use the results of ref. \cite{6} 
and to make a comparison with their results.) 
The linear equations determining the appropriate value 
of $n$ and the new values of $\alpha_{c}$, $B$, and $Z_{0}$ were 
specified by demanding that the model masses of the proton,
neutron, $\Delta$, and $\Omega^-$, be equal to the 
empirical masses after the 
inclusion of the effective instanton-induced interaction. 
We will call the approach of 
ref. \cite{2} the linearized approximation. 

In this paper we go beyond this approximation, performing a 
refitting of baryon masses which allows their radii to vary. 
It turns out that it leads to larger instanton densities
allowed inside the MIT bags, and correspondingly to a stronger 
share of instantons in the energy balance of the baryon bags, 
accompanied by decreased $\alpha_c$, as well as by acceptable, 
and for nucleons even highly favorable \cite{GerryEtAl,2}, changes 
in the baryon radii.      The most important for the present paper 
is amending the results on the nucleonic scalar strangeness 
obtained in ref. \cite{Klabucar:1998pb}.  On the one hand, larger 
instanton densities now lead to increased contributions of the 
instanton--induced interaction to the total scalar strangeness of 
the nucleon. On the other hand, the basic MIT bag strangeness 
(\ref{naiveNs}) found by Donoghue and Nappi \cite{Donoghue:1985bu}, 
if not far from its naive limit, may well still represent the main 
contribution. If it does, the total nucleon strangeness decreases 
with diminishing bag radii, which are in turn associated with 
growing instanton densities.

\section{Refitting of the baryon bag parameters}
\label{Refitting}

Except for removing the linearized approximation, 
{\it i.e.}, replacing it with the refitting where the bag radii are 
not frozen any longer, the incorporation of the instanton effects 
in the MIT bag follows ref. \cite{2} closely. The same holds also
for other model details, such as the fixed model inputs,
the nonstrange and strange quark mass parameters ($m_u = m_d = 0$ 
and $m_s = 279$ MeV, respectively) and quark-antiquark ($q{\bar q}$)
condensate $\langle 0 | {\bar q} q | 0 \rangle = - (240 \, {\rm MeV})^3$.  
Thus, to keep the present paper as concise as possible, we refer to 
ref. \cite{2} for all model details and parameters, 
and to ref. \cite{Klabucar:1998pb} 
for the corresponding strangeness calculation. (For detailed 
technicalities of the latter, ref. \cite{Klabucar:2000gu} may 
also be found helpful.)

Here we just recall that the effective instanton-induced interaction 
${\cal L}_{I}$,
causing the instanton-induced mass shift $E^{\cal B}_{I}$ of the
baryon $\cal B$, 
is the sum of the one-, two-, and three-body terms, denoted by 
${\cal L}^{I}_1$, ${\cal L}^{I}_{2}$, and ${\cal L}^I_3$, respectively:

\begin{equation}
E^{\cal B}_{I} = \langle {\cal B} |: - {\cal L}_{I} :| {\cal B} \rangle
= \langle {\cal B}|: - {\cal L}^{I}_{1}
- {\cal L}^{I}_{2} - {\cal L}^{I}_{3}:|{\cal B} \rangle  \, ,  
\label{L_I}
\end{equation}

\noindent and is defined in detail in ref. \cite{2}. 
The explicit expressions 
for the one- and two-body contributions ($\Delta M_{\cal B}^{(1)}$ and 
$\Delta M_{\cal B}^{(2)}$, respectively) are also given in ref. \cite{2}.

Before proceeding, let us make two comments regarding our choice 
of the instanton-induced interaction ${\cal L}_{I}$. It was derived by 
Nowak et al. \cite{4} in the framework of random instanton liquid model
(RILM).
They arrived at the interaction corresponding to the well-known one of 
Shifman, Vainshtein, and Zakharov (SVZ) \cite{e}, apart from the effects 
of smearing over the size of an instanton. In the limit of no smearing,
it reproduces our chosen \cite{2,Klabucar:1998pb,Klabucar:2000gu} 
local ${\cal L}_{I}$, which is essentially 
the same as the SVZ interaction \cite{e}. Since the SVZ interaction 
is induced by a single (anti-)instanton, our modeling misses 
multi-instanton effects. Their importance, however, was stressed in, 
{\it e.g.}, refs. \cite{9,Schafer:1996wv}, putting in doubt the validity 
of the single-instanton approximation. The caveat is that these 
effects can be important when baryon bags have diameters larger than 
average separation of (anti-)instantons, and this will turn out to be
%which, as it turns out, is 
the presently relevant situation (since we will find instanton densities 
inside bags up to one third of the QCD vacuum value of 1 fm$^{-4}$).
Nevertheless, as discussed especially
in ref. \cite{2}, we should recall that this interaction was introduced
and used \cite{2,Klabucar:1998pb,Klabucar:2000gu} with the aim of
capturing the intermediate-range ($\sim \frac{1}{3}$ fm) QCD effects, 
and the interaction we adopted is suitable for that, since the 
average instanton size is $\rho \approx \frac{1}{3}$ fm 
\cite{5,8,9,4,Faccioli:2001ug}. 
Hopefully, it may capture the effects at ranges even a little 
beyond $\frac{1}{3}$ fm, since Nowak et al. \cite{4}
took into account the delocalization of zero modes{\footnote{Thus,
Nowak et al. \cite{4} took into account the insights of, {\it e.g.},
refs. \cite{9}, concerning the importance of summing up a large 
number of interactions with different instantons. The review 
by Schafer and Shuryak still points out as useful their results 
and RILM approach in general, observing that interactions among
instantons (and hence their correlations) are important but not
dominant \cite{Schafer:1996wv}.}}. In keeping with the basic idea 
of the MIT model, one assumes that really long range ({\it i.e.}, 
confinement) effects are modeled well by the confining bag boundary.

The second comment is devoted to clarifying our inclusion of the 
one-body term ${\cal L}^{I}_{1}$ into the bag model calculations of the 
instanton-induced contribution (\ref{L_I}) to baryon masses. The term
${\cal L}^{I}_{1}$ has in fact the form of a mass term, and can be thought
of as the self-energy, or the effective mass that a quark acquires from
the effective interaction caused by the instanton liquid through which 
quarks move. 
Now imagine that we are working in some kind of constituent quark model 
where one from the start uses effective constituent quark masses to 
parameterize ``dressing" by nonperturbative QCD. The self-mass part of 
the instanton effects would in that case be already included in the 
constituent mass parameters. Using ${\cal L}^{I}_{1}$ in the 
baryon mass calculation would therefore be double-counting, so
in that case it must be dropped from Eq. (\ref{L_I}). On the other hand, 
if we employ some approach where one uses the current, Lagrangian quark 
masses, like in the MIT bag model used presently and in refs.
\cite{2,Klabucar:1998pb,Klabucar:2000gu}, ${\cal L}^{I}_{1}$ should be 
included in the calculation on equal footing with ${\cal L}^{I}_{2}$ 
and ${\cal L}^{I}_{3}$. This procedure was criticized by Dorokhov
\cite{Dorokhov:1994fv} on the grounds that in the bag model, the 
role of the quark constituent mass is played by the single-quark 
kinetic energy eigenvalue resulting from the boundary condition 
confining the quarks inside the bag. According to this view, the 
quark nonperturbative dressing due to the ${\cal L}^{I}_{1}$ part 
of the instanton-induced interaction would already be taken into 
account by the linear bag boundary condition. However, we do not 
accept this view because this boundary condition serves to incorporate
confinement, prohibiting quark separations larger than the bag diameter 
scale of the order of some 2 fm, while instantons are {\it not} 
responsible for confinement \cite{Greensite:1984sb,Simonov:1989tb} 
(contrary to what was thought in early days of instanton physics). 
Admittedly, this argument is so far only qualitative in the sense that 
in the model context it is not possible to delineate 
precisely beyond which scale confinement effects overwhelm instanton 
effects. Nevertheless, the argument becomes stronger and more precise
if one remembers the discussion in the previous passage: there, it was 
noted that the adopted instanton-induced interaction approximates 
well the nonperturbative QCD effects at intermediate ranges around 
$\frac{1}{3}$ fm, but not much further than that, and certainly 
not up to confinement scales of the order of the bag diameter.

As remarked above, the interaction ${\cal L}_I$ is actually the same as the 
well-known SVZ interaction \cite{e}, including the (only seemingly different 
\cite{4,f}) three-body term ${\cal L}^{I}_{3}$. This term was in fact 
discussed in ref. \cite{2} because, at that point, it was not clear whether 
the contribution of ${\cal L}^{I}_{3}$ vanished for $\Lambda$,
as it did for other baryons. Therefore, 
ref. \cite{2} avoided the need to compute the contribution 
from the complicated-looking ${\cal  L}^{I}_{3}$ by 
showing that it could contribute only to the mass of the $\Lambda$ 
and by omitting the $\Lambda$ from the analysis. 
However, it turns out that the mass shift due to the 
three-body interaction, if nonzero, must be 
small for the $\Lambda$ \cite{Duck}.
(In explicit evaluation one can see that all terms in 
the ${\cal L}^{I}_{3}$-contribution would cancel in the 
SU(3)-symmetric limit. 
This contribution slightly differs from 
zero only because the strange-quark wave functions differ 
somewhat from the nonstrange ones.)
Neglecting therefore this contribution to $M_\Lambda$,
the total instanton-induced mass shift (\ref{L_I}) 
consists of one- and two-body contributions only \cite{2}:

\begin{equation}
E^{\cal B}_{I} = \langle {\cal B}|: - {\cal L}^{I}_{1}  
- {\cal L}^{I}_{2}: |{\cal B} \rangle \equiv
 \Delta M^{(1)}_{\cal B} + \Delta M^{(2)}_{\cal B}, 
\label{E_I}
\end{equation}

\noindent for all baryons ${\cal B}$, including the $\Lambda$. 
There is hence no need to drop the $\Lambda$
from the analysis, so in this respect, this calculation is slightly
more complete than in ref. \cite{2}. 
However, when we did drop the $\Lambda$ from
the present fit in order to check the effects thereof, 
the results were affected very little.

Therefore, the only significant difference in modeling with respect to 
ref. \cite{2} is that in the present paper we want to perform 
a full refitting of the model parameters, including the variation 
of the bag radii. Maybe some reader might then object that for 
each baryon $\cal B$, its bag radius would become a new free 
parameter, and the number of fitting parameters would become larger 
than the number of experimental baryon masses $M^{\cal B}_{exp}$ to 
be be fitted. Fortunately, this is not so, because each radius 
$R_{\cal B}$ of a bag in equilibrium must satisfy the 
pressure-balance condition. That is, the equilibrium bag radius 
$R_{\cal B}$ of the baryon $\cal B$ is fixed by minimizing the 
bag model mass $M^{\cal B}_{bag}$,
\begin{equation}
\frac{d M^{\cal B}_{bag}}{d R_{\cal B}} = 0  \, \qquad
({\cal B} =  N, \Lambda, \Sigma, \Xi, \Delta, \Sigma^*, \Xi^*, \Omega)
\, \,  ,
\label{equilRcond}
\end{equation}
and is not a free, adjustable parameter 
like $Z_{0}, {\alpha}_{c},B$ and $n$.

The MIT bag energy functional $M^{\cal B}_{bag}$ of the baryon ${\cal B}$ 
now depends also on the instanton density $n$, because 

\begin{equation}
M^{\cal B}_{bag} [R_{\cal B},Z_{0}, {\alpha}_{c},B, n] = E^{\cal B}_{Q} + E^{\cal B}_{0} + E^{\cal B}_{M}
+ E^{\cal B}_{E}+ E^{\cal B}_{V}+ E^{\cal B}_{I}
\label{M_B}
\end{equation}

\noindent
now contains the instanton contribution $E^{\cal B}_{I}$
(\ref{L_I}),
in addition to the kinetic energy of the confined quarks 
$E^{\cal B}_{Q}$, the zero-point energy 
$E^{\cal B}_{0}$, the color magnetic energy $E^{\cal B}_{M}$,
the color electric energy $E^{\cal B}_{E}$, 
and the volume energy $E^{\cal B}_{V}$. The expressions for these 
five latter contributions are given in ref. \cite{6}.
(In Eq. (\ref{M_B}), the dependence of $M^{\cal B}_{bag}$ on 
the quark mass parameters $m_u, m_d, m_s$ and the condensate 
$\langle 0 | {\bar q} q | 0 \rangle$ is not indicated, as they
are not adjustable parameters but fixed model inputs.)

In the circumstances explained above, the most practical and
numerically tractable way to perform the model fit to the 
empirical baryon masses, is to pose it as the problem of
minimization of the positive definite functional $F$
 
\begin{equation}
F[\{R_{\cal B}\}, Z_{0}, {\alpha}_{c}, B, n] \equiv
F_{M} + F_{R}  \, ,
\label{F}
\end{equation}
\begin{equation}
F_{M} \equiv 
\sum_{{\cal B} } (M^{\cal B}_{exp} - M^{\cal B}_{bag})^{2} \, ,
\label{FM}
\end{equation}
\begin{equation}
F_{R} \equiv \sum_{{\cal B} } \frac{1}{{\cal M}^2} 
\left(\frac{d M^{\cal B}_{bag}}{d R_{\cal B}}\right)^2 \, ,
\label{FR}
\end{equation}
where the both sums run over the baryons $\cal B$ in the ground-state 
octet ($N, \Lambda, \Sigma, \Xi$) and decuplet ($\Delta, \Sigma^*, \Xi^*, 
\Omega$). Thus, note that in the present fitting procedure 
all baryon masses enter on an equal footing, whereas ref. \cite{2},
similarly to ref. \cite{6}, chooses some masses somewhat arbitrarily 
to fix the parameters and predict the other masses. 

In the functional $F$, the first sum, $F_{M}$, represents the deviation 
of the bag model masses $M^{\cal B}_{bag}$ from the experimental baryon 
masses $M^{\cal B}_{exp}$. The second sum, $F_R$, is a measure of the 
deviation from the situation of the perfectly satisfied pressure-balance 
condition. The role of the constant ${\cal M}$ is just to ensure that the 
both terms have the same dimension, and we choose the typical baryonic 
mass scale of 1 GeV to fix its value: ${\cal M} = 1$ GeV.  
(Of course, there is some arbitrariness in the choice of the functional 
$F$; for example, we could replace $\cal M$ by $M^{\cal B}_{exp}$ in each 
term of the sum $F_R$. However, we have checked that varying the scale 
$\cal M$  does not influence our results significantly, giving us 
confidence that this arbitrariness is not a problem in practice.) 

The functionals defined by Eqs. (\ref{F})-(\ref{FR}), namely $F$, $F_{M}$ 
and $F_R$, all depend 
on the model parameters $Z_{0}, {\alpha}_{c}, B, n$ and on the set of the
bag radii $\{R_{\cal B}\}$ of the octet and decuplet baryons. The strict 
approach to the model fitting through the functional minimization would be 
to pick the initial values of the free parameters, 
$Z_{0}^{(0)}, {\alpha}_{c}^{(0)}, B^{(0)}, n^{(0)}$, 
and find the equilibrium radii $R_{\cal B}$ by minimizing the
functional $F_R$ to, ideally, $F_R=0$, where the conditions 
(\ref{equilRcond}) would be strictly satisfied. Then, the functional 
$F_{M}$ should be calculated. This two-step process should be repeated 
with varied values of the free parameters over and over again by some 
minimization routine (for example based on simplex minimization) till 
$F_{M}$ is as close to zero as possible.
However, this two-step process, where $F_R$ would be minimized before 
each call to $F_{M}$, is computationally rather intractable 
in practice. 
Fortunately, it turns out that for the degree of accuracy that is sensible 
to demand from the MIT bag model (set by $F_{M}$ of the original 
fit \cite{6}), 
it is sufficient to perform the refitting 
by varying simultaneously $Z_{0}, {\alpha}_{c}, B, n$ and the set 
$\{R_{\cal B}\}$ to minimize the joint functional $F$. Nevertheless, one
should accept only those minimizations where $F_{M}$ is the overwhelming
share, and $F_R$ only a small part of $F=F_{M} + F_R$; 
otherwise the fit 
to the experimental masses would be done away from the equilibrium bag 
radii. Ideally, the aim would be $F=0$, but 
since it is not possible to model all experimental masses {\it exactly}, 
one should look for such parameter values for which $F$ is 
sufficiently small. In the present model it is sensible to demand 
$F < 3 \times 10^{-3}$ GeV$^2$, since the original MIT bag fit \cite{6}
gives $F_{M} = 3.2 \times 10^{-3}$ GeV$^2$.

The minimization of $F$ by the simplex method \cite{NAG}, which 
had already been proved as robust and reliable in earlier 
applications \cite{omega,Eilam:1986tg,Klabucar:zy,Klabucar:fe}, 
has turned out to be very suitable also in the present case.

\section{Results and discussion}
\label{Results}

It is necessary to give some thought as to which outputs of the
minimization procedure can be accepted as solutions to our problem.
The present situation is different than in ref. \cite{2}, where the
frozen radius approximation reduced the problem to solving a set 
of linear equations, so that the solution was unique once we chose 
which baryon masses would be used to fix the parameters. 
In the present case, the functional minimization finds many
local minima of the functional (\ref{F}). In which of them
the minimization will end up, depends on which part of the 
parameter space one starts from. Moreover, many of these
minima can be acceptable in the sense of sufficiently small 
value of the minimized functional $F$. We thus face the problem
of non-uniqueness of the solutions. Fortunately, the smallness of 
the minimized functional $F$ is not the only criterion; clearly, a 
fit resulting in a good mass spectrum would anyway be unacceptable 
if it also resulted in physically unacceptable values of the bag 
radii or fitting parameters. This must always be kept in mind, as the 
problem was mathematically posed in such a way that it is possible to 
get an excellent fit to the masses, but with the bag radii and 
parameters devoid of any physical justification. 

\subsection{Practically instantonless bag} 
\label{Instantonless}
The first thing to check is the limit of the vanishing instanton
density $n$ inside the bag. This is basically the case of the pure 
MIT bag model except that our model fitting is done by minimizing the
functional (\ref{F}). This is different from the original MIT bag 
fitting procedure \cite{6} where the parameters $B, Z_0$ and $\alpha_c$ 
were fixed by singling out three hadrons and constraining their model 
masses to be the experimental ones. From the model standpoint, it is 
very satisfying that for $n=0$ inside the bag, our different fitting 
procedure leads to the description of baryons very similar to the 
original MIT bag fitting procedure \cite{6}. 
In addition to that, we note that when we depart from the limit of 
vanishing instanton density inside the bag and finally allow $n\neq 0$, 
the minimization of the functional (\ref{F}) leads, among various 
outcomes, also to several solutions where the values of $n$ are 
nonvanishing but extremely small, $n \lsim 10^{-6}$ GeV$^4$. 
This is practically negligible in comparison with the 
density $n_0\approx 1.6 \times 10^{-3}$ GeV$^4$ = 1 fm$^{-4}$ 
estimated reliably ({\it e.g.}, see refs. \cite{5,8,9,4,Faccioli:2001ug})
for the nonperturbative QCD vacuum outside the bag. 
The resulting baryon masses and radii (as well as values of the 
variable model parameters) are very close to each other 
in all these cases of very small $n$, and 
also very similar to the pure MIT case ($n=0$), as one would expect,
although all those cases are formally different solutions.  It is thus 
clear that they all describe very similarly (``practically uniquely") 
the situations when $n\to 0$. This shows that non-uniqueness of 
the solutions is not a problem in practice, and the same happens 
in the more interesting cases with significant values of $n$,
discussed in the next subsection.   

\subsection{Appreciable instanton density inside the bag}
\label{Appreciable}

Let us now discuss the first major interest of this paper: the cases when 
instanton densities $n$ inside the bag are significantly different 
from zero. Indeed, in most cases we obtained interesting solutions 
where densities $n$ inside the bag are 
an order of magnitude larger than in the 
linearized approximation \cite{2}, where{\footnote{The dimensionless 
density $\widetilde n$ used in \cite{2} and $n$ are related by 
$n \equiv \widetilde{n} \rho^{-4}$, where $\rho$ is the average 
instanton radius. Throughout, we have adopted the standard value 
$\rho=1/600$ MeV$^{-1} \approx 1/3$ fm 
({\it e.g.}, see refs. \cite{5,8,9,4,Faccioli:2001ug}).}}
$n=0.266 \cdot 10^{-4}$ GeV$^4$.  
However, for all acceptable minimizations of the functional (\ref{F}),
we find they are still appreciably lower 
(at least by the factor of 3 or more)
than the nonperturbative vacuum 
density $n_0\approx 1$ fm$^{-4} = 1.6 \cdot 10^{-3}$ GeV$^4$. Therefore, 
we do not get a description of baryons which would be drastically 
different from the original MIT bag one \cite{6}, but we do obtain the 
desirable decrease of $\alpha_c$ which is noticeably stronger than the 
corresponding decrease obtained earlier in the linearized approximation 
\cite{2}. (In the case depicted in Table 1, $\alpha_c$ is by 30\%
smaller than in ref. \cite{6}.)
In the solutions with decreased $\alpha_c$, we also observe 
the decrease of the baryon radii $\{R_{\cal B}\}$. As mentioned above,
this is very desirable from the standpoint of nuclear physics, as 
explained by, {\it e.g.}, Brown {\it et al.} \cite{GerryEtAl}. 
Namely, standard 
nuclear physics descriptions favor the picture of nuclei as made of 
{\it independent} nucleons interacting by effective boson exchange, 
but the empirical sizes of nuclei indicate that the ``standard''
MIT nucleon bags with $R_N \approx 1$ fm  are already somewhat
too large \cite{GerryEtAl} for that. 
For this reason, we give in Table 1 the case with the smallest 
nucleon radius for which we managed to achieve an acceptable fit.
Other physically acceptable solutions have somewhat smaller $n$ and
somewhat larger radii.
Table 2 gives a kind of condensed overview of several representative fits;
{\it e.g.}, the last line in Table 2 summarizes Table 1, the case with 
the highest $n$ which leads to a fit acceptable by all criteria.
The general features of the acceptable fits are the following:

{\it a)} 
The values of the functional $F$ are around 1.3 to $1.2 \times 
10^{-3}$ GeV$^2$ (out of which only less than a percent is $F_R$). 
This gives the rough limit on the accuracy of reproduction of the 
mass spectrum within the present model. The average deviation from
an experimental baryon mass is 11 MeV. In fact, the predictions for
the masses of $N$ and $\Sigma$ are the worst. They are too high for 
$N$ and too low for $\Sigma$ by some twenty MeV. The other masses are 
within 10 MeV from the experimental masses. (In the MIT fit \cite{6},
the $N$ mass belongs to those constrained to experimental values to
fix the model parameters, but then the $\Sigma$ mass is too low by
45 MeV.) 
Overall, our approach to fitting of baryon masses gives noticeably
smaller sum of squared deviations from the empirical baryon masses,
$F_{M}$, than the original MIT bag fit \cite{6} and the linearized 
approximation (where $F_{M} = 3.5 \times 10^{-3}$ GeV$^2$) \cite{2}.

{\it b)}  Going beyond the linearized approximation and thereby 
allowing the bag radii to vary, leads to some significant changes 
with respect to the results in linearized approximation. Notably, 
Table 1 shows the instanton contributions to baryon energies are an 
order of magnitude larger than in the linearized approximation \cite{2}.
Such instanton contributions are present not only in Table 1, but in 
large majority of fits, since the instanton densities in most 
of the presently obtained solutions are an order of magnitude larger 
than $n$ obtained in the linearized approximation \cite{2}. 
Nevertheless, since the instanton contributions to bag masses are still 
much smaller than other contributions (except $E_E^{\cal B}$), the general
picture of baryons is not {\it drastically} altered with respect to the 
original MIT bag phenomenology \cite{6}.  

{\it c)}  In most cases this relatively large $n$ inside the bag 
leads to the decrease of $\alpha_c$, although there are also some fits 
with relatively large $n$ where $\alpha_c$ grows back close to 
its MIT value \cite{6}.  Then, however, such a larger $\alpha_c$ 
is also accompanied by an excessive increase of bag radii.
In particular, this yields a nucleon radius even larger than in the 
MIT case \cite{6}, so that such solutions must be discarded as 
unacceptable from the point of view of nuclear physics as explained 
above. 
The interdependence of the model parameters and the baryon
bag radii which minimize $F$ is such that $\alpha_c$ decreases
while $Z_0$ and $B$ increase with the decrease of the bag radii.
This is illustrated in Table 2, which,
for four different fits, displays the {\it average}
octet ($O$) and decuplet ($D$) radii, ${\overline R}^O$ and 
${\overline R}^D$, for four different fits.  The notion of the 
average multiplet radii ${\overline R}^O$ and ${\overline R}^D$
is useful since the octet baryon radii are similar to each other,
and the decuplet baryon radii are similar to each other.
The decuplet radii are also some 10 \% larger than the radii of the 
octet baryons.

\subsection{Instanton-induced strangeness inside the bag}
\label{Strangeness}

An inspection of ref. \cite{Klabucar:1998pb} easily
shows that going beyond the linearized approximation, and
the above effects thereof, do not change the results of
ref. \cite{Klabucar:1998pb} on 
the vector, axial-vector and pseudo-scalar strangeness 
of the nucleon bag: the instanton-induced 
contributions to them are still vanishing.

In contrast to that, the instanton-induced 
scalar strangeness is enhanced 
an order of magnitude over what it was in the 
linearized approximation \cite{2}, following the
increase of the instanton density $n$. 
This is seen in Table 3, which shows the dependence
on the instanton density, or on the bag radius associated 
with this density, of various scalar strangeness 
components of the nucleon. 
The instanton-induced contributions due to 
${\cal L}^{I}_{1}$ and ${\cal L}^{I}_{2}$, respectively
denoted by $\langle N| {\bar s} s |N  \rangle_{{\cal L}^I_1}$
and $\langle N| {\bar s} s|N \rangle_{{\cal L}^I_2}$, 
comprise the overwhelming share of the total
instanton-induced contribution 
$\langle N| {\bar s} s |N \rangle_{{\cal L}_I}$.
We do not display
$\langle N| {\bar s} s |N \rangle_{{\cal L}^I_3}$,
the contribution due to ${\cal L}^I_3$, as it
contributes only to the third decimal place.

Although our present interest are the instanton-induced 
contributions, we should also comment on the basic strangeness 
of the nucleon MIT bag, $\langle N| {\bar s} s |N \rangle_{\rm basic}$
(found by Donoghue and Nappi \cite{Donoghue:1985bu}),
\begin{equation}
\langle N| {\bar s} s |N \rangle_{\rm basic} \equiv (\eta - 1) 
\langle 0 | {\bar q} q | 0 \rangle \, \frac{4\pi}{3}\, R_N^3 \, .
\label{naiveNs}
\end{equation}
It is the product of the nucleon bag volume $V_N = (4\pi/3)R_N^3$
and $\langle 0 | {\bar q} q | 0 \rangle$, the expectation value 
of the ${\bar q} q$ scalar condensate in the true, 
nonperturbative QCD vacuum, but also of the factor $\eta - 1$ which 
has unfortunately remained quantitatively undetermined. 
Its determination is beyond the scope of the present paper. 
Let us just quote \cite{Donoghue:1985bu} that the 
$\eta$ ($0 < \eta < 1$) is in general some decreasing function 
of the bag radius, since $R_N \to \infty$ corresponds to $\eta \to 0$.
The case $\eta = 0$ is called the naive bag model limit and obviously 
maximizes the basic bag strangeness (\ref{naiveNs}).  This limit was, 
for definiteness, the only case of the basic bag strangeness 
$\langle N| {\bar s} s |N \rangle_{\rm basic}$ considered in ref. \cite{Klabucar:1998pb}.
Although in the present paper even the $\eta = 0$ limit of
$\langle N| {\bar s} s |N \rangle_{\rm basic}$ is not so much larger
than (now increased) $\langle N| {\bar s} s |N \rangle_{{\cal L}_I}$
as was the case in ref. \cite{Klabucar:1998pb}, it
is still larger by an order of magnitude for all radii displayed in Table 3.
The most widely accepted value of the condensate, 
adopted also in ref. \cite{2} and the present paper,
$\langle 0 | {\bar q} q | 0 \rangle = - (240 \, {\rm MeV})^3$,
leads to $\langle N| {\bar s} s |N \rangle_{\rm basic}$ exceeding 
considerably the empirical value of the total scalar strangeness
(determined by, {\it e.g.}, ref. \cite{Zhitnitsky:1996ng}, 
from the $\sigma$-term estimated from the $\pi N$ scattering 
data and the masses of $\Xi, \Sigma$ and $\Lambda$),
\begin{equation}
\langle N| {\bar s} s |N \rangle \approx 2.8 \, .
\label{EXPsbars}
\end{equation}
Of course, lower values of $\langle 0 | {\bar q} q | 0 \rangle$ trivially
decrease $\langle N| {\bar s} s |N \rangle_{\rm basic}$. For example, the
choice $\langle 0 | {\bar q} q | 0 \rangle = - (200 \, {\rm MeV})^3$,
as in ref. \cite{Klabucar:1998pb}, amounts to reducing
values of $\langle N| {\bar s} s |N \rangle_{\rm basic}/(1-\eta)$ in Table 3
by the factor $(200/240)^3 = 1/1.728$, but this still gives 
rather large values.
Instantons inside the bag help with that. Admittedly,
Eq. (\ref{naiveNs}) shows clearly that 
$\langle N| {\bar s} s |N \rangle_{\rm basic}$ is not directly dependent 
on instantons and their density $n$ inside the bag, but there is
an indirect connection: first, the volume factor in
Eq. (\ref{naiveNs}) decreases with radii as $R_N^3$, and 
diminishing radii are associated with increasing $n$. 
Second, $1-\eta$ {\it also} falls with $R_N$. Thus, 
even if $\langle N| {\bar s} s |N \rangle_{\rm basic}$ in the original, 
instantonless MIT bag model would be too close to its (too large) 
naive value, this potential problem would now be alleviated 
(more strongly than $R_N^3$)
by lower values of $R_N$, occurring at higher $n$.

As already stressed, the most interesting effect of
presently increased values of the instanton density $n$ is 
the considerable enhancement of the instanton-induced scalar 
strangeness, and we would like to point out that this enhancement 
is {\it not} due to a favorable choice of fixed input parameters 
$m_u , m_d, m_s$ and $\langle 0 | {\bar q} q | 0 \rangle$.
In fact, our adoption of the fixed input parameters of refs. \cite{2} 
and \cite{6} was motivated not only by easiness of comparison
with these papers. This choice is also suitable for
stressing that the present enhancement (of the
{\it instanton-induced} strangeness) is not an effect
of the choice of the model parameters. This is because
the values of the quark mass parameters and of the vacuum 
quark-antiquark (${\bar q} q$) scalar condensate used in 
ref. \cite{2} and in obtaining all presently displayed results
($m_u = m_d = 0$, $m_s = 279$ MeV, 
$\langle 0 | {\bar q} q | 0 \rangle = - (240 \, {\rm MeV})^3$),
actually lead to smaller instanton-induced scalar strangeness than 
those used in ref. \cite{Klabucar:1998pb}
($m_u = m_d = 8$ MeV, $m_s = 200$ MeV,
$\langle 0 | {\bar q} q | 0 \rangle = - (200 \, {\rm MeV})^3$).
That the latter choice \cite{Klabucar:1998pb} of these inputs
gives (at a given instanton density $n$) even more enhanced 
instanton-induced scalar strangeness than that in Table 3,
is most easily understood if one notes the role of 
the characteristic pre-factors, denoted by ${\cal F}_{f}$'s 
in refs. \cite{2,Klabucar:1998pb}, appearing in the 
instanton-induced interaction ${\cal L}_I$. 

The factor ${\cal F}_{f}$ pertaining to a flavor $f$ $(f=u,d,s)$, 
is composed of the corresponding quark mass parameter $m_{f}$, 
the average instanton size $\rho \simeq \frac{1}{3} \,{\rm fm}$ 
\cite{5,8,9,4,Faccioli:2001ug}, and the ${\bar q} q$ condensate
$\langle 0|\overline{q} q|0\rangle$, in the following way:
\begin{equation}
{\cal F}_{f} \equiv 
\frac{1}{m_{f}\rho - \frac{2\pi^{2}}{3} \rho^{3} 
\langle 0| \overline{q} q|0\rangle} \, , 
\, (f=u,d,s).
\label{Ff}
\end{equation}
Obviously, smaller values of $m_u, m_d, m_s$ and 
$\langle 0| \overline{q} q|0\rangle$ will increase 
${\cal F}_{f}$'s, and vice versa.

Let us consider the concrete sets of input parameters, those of 
refs. \cite{2} and \cite{Klabucar:1998pb}. Changing $m_u = m_d$ 
from 0 to 8 MeV actually does not have significant influence on 
the instanton-induced strangeness, since $m_u$ and $m_d$ are 
anyway small at the hadronic mass scale 
(where 8 MeV can be approximated by 0).
Nevertheless, the decrease of $m_s$ from 279 MeV to 200 MeV 
is quite important for further increasing the instanton-induced 
strangeness significantly over the values in Table 3. In fact, 
the effect thereof is comparable to the effect of the decrease of 
$| \langle 0| \overline{q} q|0\rangle |$ from
$(240 \, {\rm MeV})^3$ to $(200 \, {\rm MeV})^3$.

Besides the effect on ${\cal F}_{f}$'s, the change of quark masses 
changes the quark wave functions, and, more importantly, the quark 
energy denominators appearing in the calculation of the nucleon 
strangeness (as can be seen in ref. \cite{Klabucar:1998pb}).
This way, the decrease of $m_s$ from 279 MeV to 200 MeV increases 
still further the instanton-induced strangeness.
(Again, the increase of $m_u = m_d$ from 0 to 8 MeV is too small
to have a significant influence.)
The effect of the quark energy denominators and wave functions
is not so clearly disentangled as the effect of the 
${\cal F}_{f}$--factors, so the explicit calculation is needed 
to show that the effect is of comparable magnitude. However, 
the important thing for the present discussion is that this 
effect changes the nucleon strangeness in the same direction
as the ${\cal F}_{f}$--factors.

\section{Conclusion}
\label{Conclusion}

To summarize, we first remark that we did not perform a
``first-principle"-type
calculation of the probability of penetration of the instanton liquid
from the surrounding nonperturbative QCD vacuum into the bag.
Rather, we performed model fits to baryon masses and these fits
showed which values of the instanton densities can be accommodated 
inside the MIT bag (in a physically acceptable way)
and what the effects thereof would be. 
In the present paper, we went beyond the linearized approximation of 
ref. \cite{2}, and the bag radii were allowed to vary in the course
of parameter fitting, which was performed so that the radii had to 
satisfy the pressure-balance condition.
In this approach, the importance of the instanton-induced interaction
allowed to act inside the quark bag is increased in every way, and not
only for the baryon mass shifts, the size of which follows the increase
of the instanton density inside the bag. We namely found that 
the MIT bag interior can accommodate instanton densities an order of 
magnitude larger than found in the linearized approximation \cite{2}. 
They grow faster than the inverse of the bag volume with decrease of 
the bag radii returned by the fitting procedure. The growth of the 
instanton-induced scalar strangeness of the nucleon is even slightly 
faster than that when the nucleon radius falls. The instanton-induced
part of the nucleon strangeness is now an order of magnitude larger 
than the instanton-induced strangeness found in the linearized 
approximation \cite{Klabucar:1998pb}. 
The quantity (\ref{naiveNs}), considered as the basic MIT bag 
contribution to nucleon strangeness \cite{Donoghue:1985bu}, remains 
undetermined also in the present work, but we could show that it 
must be smaller in the instanton-enriched MIT bag than in the 
original MIT bag. This is good, because this quantity alone has the
potential to overshoot strongly the empirical value (\ref{EXPsbars})
of $\langle N| {\bar s} s |N \rangle$.
Also, it turns out that allowing for the possibility of 
%$n \neq 0$ 
instanton densities significantly different from zero inside the MIT bags now
enables the favorable, up to 30~\% decrease of $\alpha_c$. More importantly, 
it also enables the decrease of the nucleon MIT bag radius by more than 10 \%, 
improving somewhat the consistency of the MIT bag model with nuclear physics.

After so summarizing the concrete results of this paper, we close
by making a more speculative comment on how our enabling substantial 
instanton-liquid densities inside the MIT bag seems to improve the 
consistency of the model. Let us first note that, contrary to some
earlier statements \cite{Witten:1978bc}, 
Sch\" afer \cite{Schafer:2002af,Schafer:2002pb} has 
recently shown that the instanton liquid model, including the one 
we use, is not necessarily in conflict 
with the expansion in large $N_c$, the number of QCD colors. Then
we recall an observation of Bardeen and Zakharov \cite{Bardeen:1979nu}
concerning the $N_c$ scaling of the bag constant $B$.
Inside hadrons, which can be modeled by the bag model,
the quark color fields most probably suppresses instantons
and other nonperturbative fluctuations. However,
a smooth large-$N_c$ limit seems to indicate \cite{Bardeen:1979nu}
that the suppression of these fluctuations, including
instantons, is not very strong (in keeping with relatively
low values of $B$ coming from phenomenological fits). 
Our modification of the bag, containing considerable 
instanton-liquid densities inside, is obviously more consistent 
with the Bardeen and Zakharov's result \cite{Bardeen:1979nu}
than is the original MIT model, where this suppression is complete.

%%%%%%%%%%%%%%%%%%%%%%%%%%%%%%%%%%%%%%%%%%%%%%%%%%%%%%%%%%%%%%%%%%%

\section*{Acknowledgments}
\noindent 
The support of the Croatian Ministry of Science and
Technology contract 0119261 is gratefully acknowledged.

\newpage

\vspace{6mm}

\vspace{5mm}
\begin{center}
\begin{tabular}{|c|c|c|c|c|c|c|c|c|c|}
\hline
&&&&&&&&& \\
Baryon ${\cal B}$ & $\, \, M^{\cal B}_{exp}\,\,  $ &  
$\, \,  M^{\cal B}_{bag}\, \, $ &$\,\,\,\, R_{\cal B}\,\,\, \,$  &  
$\, \,\, E_{0}^{\cal B}\, \,\, $ & $\,\,\, E_{V}^{\cal B}\,\,\, $ & 
$\,\,\, E_{Q}^{\cal B}\,\,\,$ & $\,\,\, E_{M}^{\cal B}\,\,\, $ & 
$\,\,\, E_{E}^{\cal B}\,\,\, $ & $\,\, E_{I}^{\cal B}\,\, $\\
&&&&&&&&&\\
\hline
&&&&&&&&& \\
$N$&$0.938$&$0.959$&$4.365$&$-0.526$&$0.176$&$1.404$&$-0.128$&
$0.000$&$0.033$\\
&&&&&&&&& \\
$\Lambda$&$1.116$&$1.120$&$4.383$&$-0.524$&$0.178$&$1.557$&$-0.127$&
$0.003$&$0.033$\\
&&&&&&&&& \\
$\Sigma^{+}$&$1.189$&$1.172$&$4.529$&$-0.507$&$0.197$&$1.513$&$-0.094$&
$0.003$&$0.061$\\
&&&&&&&&& \\
$\Xi^{0}$&$1.315$&$1.306$&$4.475$&$-0.513$&$0.190$&$1.688$&$-0.109$&
$0.003$&$0.047$\\
&&&&&&&&& \\
$\Delta$&$1.232$&$1.248$&$5.130$&$-0.448$&$0.286$&$1.195$&$~~0.109$&
$0.000$&$0.107$\\
&&&&&&&&& \\
$\Sigma^{*}$&$1.385$&$1.388$&$5.073$&$-0.453$&$0.277$&$1.370$&$~~0.096$&
$0.003$&$0.095$\\
&&&&&&&&& \\
$\Xi^{*}$&$1.533$&$1.526$&$5.027$&$-0.457$&$0.269$&$1.543$&$~~0.084$&
$0.003$&$0.083$\\
&&&&&&&&& \\
$\Omega^{-}$&$1.672$&$1.661$&$4.978$&$-0.461$&$0.261$&$1.716$&$~~0.074$&
$0.000$&$0.071$\\                                                        
&&&&&&&&& \\
\hline
\multicolumn{10}{|c|}{ } \\
\multicolumn{10}{|c|}{   {$B^{1/4}=0.150 $ GeV}   \quad
{$Z_{0}=2.296$}  \quad { $\alpha _{c}=0.394$}
\quad  { $ n = 0.512 \cdot 10^{-3}$ GeV$^4$}  }   \\                      
\multicolumn{10}{|c|}{ } \\
\hline
\end{tabular}
\end{center}

\parbox[t]{14cm}

\vspace{1.5cm}

\noindent {\bf Table 1:}
The fit for the input quark masses $m_u = m_d = 0$, $m_s=279$ MeV
and the quark-antiquark vacuum condensate
$\langle 0 | {\bar q} q | 0 \rangle = - (240 \, {\rm MeV})^3$. 
We display the separate energies $E_X^{\cal B}$ ($X=0,V,Q,M,E,I$)
contributing to $M^{\cal B}_{bag}$, the mass of the baryon bag, 
to be compared with the corresponding experimental baryon mass 
$M^{\cal B}_{exp}$ in the first column. 
The output values of the bag model parameters $B, Z_{0}, \alpha _{c}$ 
and $n$ are given in the lowest part of Table 1.
All the masses and energies are given in GeV, and the bag radii 
$R_{\cal B}$ in inverse GeV, while $Z_{0}$ and $\alpha _{c}$ 
are dimensionless.

\newpage

\vspace{6mm}

\vspace{5mm}
\begin{center}
\begin{tabular}{|c|c|c|c|c|c|c|}
\hline
&&&&&& \\
$ n \times 10^3$  & $\alpha_c$  & $  B \times 10^4  $ &  $Z_0$ & 
$\overline R^O $  & $ \overline R^D $ & $ F \times 10^3  $\\
&&&&&&\\

[GeV$^4$]&   & [GeV$^4$] &  & 
  [GeV$^{-1}$]   &  [GeV$^{-1}$]  &  [GeV$^2$] \\
&&&&&& \\
\hline
&&&&&& \\
$0.290$&$\,\, 0.485 \,\, $&$4.031$&$\,\, 1.865\,\,$&$5.0$&$5.6$&$1.14$\\
&&&&&& \\
$0.310$&$\,\, 0.474 \,\, $&$4.188$&$\,\, 1.930\,\,$&$4.9$&$5.4$&$1.18$\\
&&&&&& \\
$0.398$&$\,\, 0.437 \,\, $&$4.612$&$\,\, 2.114\,\,$&$4.7$&$5.2$&$1.26$\\
&&&&&& \\
$0.512$&$\,\, 0.394 \,\, $&$5.058$&$\,\, 2.296\,\,$&$4.4$&$5.1$&$1.29$\\
&&&&&&\\
\hline
%\multicolumn{7}{|c|}{ } \\
%\hline
\end{tabular}
\end{center}
 
\parbox[t]{14cm}
 
\vspace{1.5cm}

\noindent {\bf Table 2:} 
Brief overview of some typical fits. (The fit given in Table 1 is 
one example of them.) The values of functional $F$ in the last column 
shows good quality of the fits. The interdependence of the adjustable 
bag parameters ($n$, $\alpha_c$, $B$, $Z_0$) and the bag radii 
is summarily depicted utilizing the average octet and decuplet 
radii, $\overline R^O$ and $\overline R^D$, respectively.

\newpage

\vspace{6mm}

\begin{tabular}{|c|c|c|c|c|c|}\hline
\parbox{8ex}{$n\times 10^3$\\[1.5ex]\ [GeV$^{4}$]}&
\parbox{8ex}{\begin{center}$R_N$\\[1.5ex] [GeV$^{-1}$]\end{center}}  &
$\langle N| \bar{s}s |N \rangle_{\calL^I_1}$ &
$\langle N| \bar{s}s |N \rangle_{\calL^I_2}$ &
 $\langle N| \bar{s}s |N \rangle_{\calL_{I}}$  &
$\frac{\displaystyle \langle N|  \bar{s}s |N \rangle_{\textrm{\scriptsize basic}}}{
\displaystyle (1-\eta)}$ \\
\hline
&&&&&\\
0.290 & 4.994  &  0.22 & 0.09 & 0.31 & 7.21  \\
&&&&&\\
0.310 & 4.909  &  0.24 & 0.10 & 0.34 & 6.85  \\
&&&&&\\
0.398 & 4.658  &  0.29 & 0.15 & 0.44 & 5.85  \\
&&&&&\\
0.512 & 4.365  &  0.36 & 0.22 & 0.58 & 4.82 \\ 
&&&&&\\\hline
\end{tabular}

\vspace{1.5cm}

\noindent {\bf Table 3:}
Dependence of the scalar strangeness of the nucleon on the instanton 
density $n$, or on the bag radius $R_N$ associated with this density. 
For comparison, we recall that in the linearized approximation 
\cite{Klabucar:1998pb} the instanton-induced strangeness from $\calL^I_1$ was
$\langle N|\bar{s}s|N\rangle_{\calL^I_1}=0.035$, whereas the contribution 
from $\calL^I_2$ was $\langle N|\bar{s}s|N \rangle_{\calL^I_2} = 0.023$
(at $R_N = 5.00$ GeV$^{-1}$$\approx 1$ fm).
In the last column, the choice $\eta=0$ maximizes the basic bag strangeness 
contribution of ref. \cite{Donoghue:1985bu}; this is nevertheless only the 
so-called naive bag model limit, and in fact $\eta$ remains undetermined.
The fixed model inputs ($m_u, m_d, m_s$ and
$\langle 0 | {\bar q} q | 0 \rangle$) are the same as in Table 1 
and Table 2, and are discussed in detail in the main text.

\end{document}